\documentclass[twocolumn,epsfig,aps]{revtex4}
\usepackage{amsmath,bm}
\usepackage{epsfig}
\usepackage{graphicx}

\begin{document}
\title{Matching Stages of Heavy Ion Collision Models}
\author{Yun Cheng$^{1,2,3}$, \footnote{Corresponding author: yun.cheng@uib.no}
L.P. Csernai$^{1,2,4}$,
V.K.~Magas$^{5}$,
B.R.~Schlei$^{6}$,
D.~Strottman$^{2,7}$}
\affiliation{
$^1$ Institute of Physics and Technology, University of Bergen,
Allegaten 55, N-5007 Bergen, Norway \\
$^2$ Frankfurt Institute for Advanced Studies,
Johann Wolfgang Goethe University,
Ruth-Moufang-Str. 1, 60438 Frankfurt am Main, Germany\\
$^3$ Institute of Particle Physics, Huazhong Normal University,
Wuhan, 430079 China\\
$^4$ MTA-KFKI, Research Inst.\ of Particle and Nuclear Physics,
1525 Budapest, Hungary\\
$^5$ Departament d'Estructura i Constituents de la Mat\`eria,
Universitat de Barcelona, Diagonal 647, 08028 Barcelona, Spain\\
$^6$ GSI Helmholtz Centre for Heavy Ion Research GmbH,
Planckstra\ss e 1, 64291 Darmstadt, Germany \\
$^7$ Dep. de Fisica Teorica and IFIC, Centro Mixto Universidad
de Valencia CSIC, Institutos de Investigaci\'on de Paterna, Aptd. 22085, 46071 Valencia,
Spain}

\begin{abstract}
Heavy ion reactions and other collective dynamical processes are frequently
described by different theoretical approaches for the different stages
of the process, like initial equilibration stage, intermediate locally
equilibrated fluid dynamical stage and final freeze-out stage.
For the last stage the
best known is the Cooper-Frye description used to generate the phase space
distribution of emitted, non-interacting, particles
from a fluid dynamical expansion/explosion, assuming a final ideal
gas distribution, or (less frequently) an out of equilibrium distribution.
In this work we do not want to replace the Cooper-Frye description, rather
clarify the ways how to use it and how to choose the parameters of the
distribution, eventually how to choose the form of the phase space
distribution used in the Cooper-Frye formula.  Moreover, the
Cooper-Frye formula is used in connection with the freeze-out problem, while
the discussion of transition between different stages of the collision
is applicable to other transitions also.

More recently hadronization and molecular dynamics models are matched
to the end of a fluid dynamical stage to describe hadronization and
freeze-out. The stages of the model description can be matched to each other
on spacetime hypersurfaces (just like through the frequently used freeze-out
hypersurface). This work presents a generalized description of how to match
the stages of the description of a reaction to each other, extending the
methodology used at freeze-out, in simple covariant form which is
easily applicable in its simplest version for most applications.
\end{abstract}
\maketitle

\section{introduction}\label{intro}
Relativistic heavy ion reactions exhibit dominant collective flow
behaviour, especially at higher energies where the number of involved
particles, including quarks and gluons, increases dramatically. At
intermediate stages approximate local equilibrium is reached, while
the initial and final stages may be far out of local equilibrium.
Also, different stages may have different forms or phases of matter,
especially when Quark Gluon Plasma (QGP) is formed.

The need to describe and match different stages of a reaction was realized
by the development of the final freeze-out (FO) description in Landau's
fluid dynamical (FD) model \cite{La53}. Then it was improved by Milekhin
\cite{Mi5861}, and a covariant simple model was given by Cooper and
Frye \cite{CF74}. In all these models the FO happened when the fluid crossed
a hypersurface in the spacetime.

At early relativistic heavy ion collisions, the initial compression and
thermal excitation was described by a compression shock in nuclear matter.
This was already pointed out by the first publications of W. Greiner and
E. Teller and their colleagues \cite{T73,G74}, and the shock took place
crossing a spacetime hypersurface (e.g. a relatively thin layer
resulting in a Mach cone). When sudden large changes happen across a
spacetime front the conservation laws and the requirement of increasing
entropy should be satisfied:
\begin{align}
&[N^{\mu}d\sigma_{\mu}]~=~0\label{nconserve}~;\\
&[T^{\mu\nu}d\sigma_{\mu}]~=~0\label{tconserve}~;\\
&[S^{\mu}d\sigma_{\mu}] ~\geq~0 \label{entropy}
\end{align}
where $ N^{\mu}=nu^{\mu} $ is the baryon current,
$S^{\mu}=su^{\mu} $ is the entropy current,
$T^{\mu\nu}$ is the energy momentum tensor, which, for a perfect fluid,
is given by
\begin{equation}
T^{\mu\nu}=(e+P)u^{\mu}u^{\nu}-Pg^{\mu\nu} \,, \label{Tmunu}
\end{equation}
where $e$ is the energy density, $P$ is the pressure, $s$ is the entropy
density, and
$n$ is the baryon density of matter. These are invariant scalars.
The $d\sigma_{\mu}$ is the normal vector of the transition hypersurface,
$u^{\mu}$ is the particle four velocity
$u^\mu = \gamma\:(1, v_x, v_y, v_z )\ = \ \gamma \:(1, \vec{v})$, normalized
to $+1$.
The square bracket means $[a]=a_1 - a_0$,
the difference of quantity $a$ over the two sides of the hypersurface.
The metric tensor is defined as
$g^{\mu \nu} = (1,-1,-1,-1)$.
We will also use the following notations: $w = e + P$,
$j = N^\mu d\sigma_\mu$
is the invariant scalar baryon current across the front,
$X=(e+P)/n^2=w/n^2$ is the generalized specific volume,
$v^2 = \vec{v}\, ^2 = v_x^2 + v_y^2 + v_z^2 $, and
$\mu,\nu...=0,1,2,3$, $i,j...= 1,2,3 = x,y,z$.

For a perfect fluid local equilibrium is assumed, thus the fluid
can be characterized by an Equation of State (EoS), $P=P(e,n)$.
Eqs. (\ref{nconserve},\ref{tconserve}) and the EoS are 6 equations,
and can determine the 6 parameters of the final state,
$e$, $n$, $P$, and $\vec{v}$.

Later Csernai \cite{Cs87,Cs94}
pointed out the importance of satisfying energy, momentum and
particle charge conservation laws across such hypersurfaces and generalized
the earlier description of Taub \cite{Ta48} to spacelike and timelike
hypersurfaces (with spacelike and timelike normals respectively).
In this situation the matter both before and after the
shock was near to thermal equilibrium, and thus the conservation laws led to
scalar equations connecting thermodynamical parameters of the two stages
of the matter: the generalized {\it Rayleigh line} and {\it Taub adiabat}
\cite{Cs87,Cs94}:
\begin{equation}
j^2=[P](d\sigma^{\mu} d\sigma_{\mu})  / [X] \ , \
[P] =[(e+P)X] / (X_1 + X_0) \ .
\label{rayligh-taub}
\end{equation}

At much higher energies, at the first stages of the collision,
the matter becomes 'transparent' and the initial state is very far from
thermal equilibrium. For this stage other models were needed to handle
the initial development, e.g. refs. \cite{Ma0102}. The initial non-equilibrium
state in this situation cannot be characterized by thermodynamical
parameters or an EoS, so the previous approach, with
the generalized {\it Rayleigh line} and {\it Taub adiabat} is not
applicable. Nevertheless, the intermediate (fluid dynamical) stage is
in equilibrium and has an EoS, while the initial state has a well defined
energy momentum tensor. In this work we will demonstrate that the
final invariant scalar, thermodynamical parameters can be determined in
this situation also from the conservation laws.

Then, Bugaev \cite{Bu96,AC99-3309}
observed that FO across hypersurfaces with spacelike normals, has problems
with negative contributions in the Cooper-Frye evaluation \cite{CF74}
of particle spectra, thus the FO
must yield an anisotropic distribution, which he could approximate with a
cut-J\"{u}ttner distribution \cite{Bu96,AC99-3309}. This is not surprising as
in the rest frame of the front (RFF) all post FO particles must move
"outwards", i.e. $p^\mu d\sigma_\mu > 0$ is required. This condition is
not satisfied by any non-interacting thermal equilibrium distribution, which
extend to infinity in all directions even if they are boosted in the RFF.
\footnote{In the following discussion we use the term anisotropic distribution
for momentum distributions in their own local rest (LR) frame. Thermal
distributions are spherical in their LR frame, although they become
anisotropic in an other frame of reference \cite{Cs94}.}

 Subsequently, another
analytic form was proposed by Csernai and Tamosiunas, the
cancelling-J\"{u}ttner
distribution \cite{Ta07}, which replaced the sharp cutoff by a continuous
cutoff, based on kinetic model results.

Parallel to this development, the FO process was analysed in kinetic, transport
approaches \cite{AL99-388,Magas:2003yp,MC0607,NW_layer,Grassi-volFO}, where
the FO happened in an outer layer of the spacetime, or in principle
it could be extended to the whole fluid (although, at early moments of
a collision/explosion, from the center of the reaction few particles
can escape).  These transport studies also indicated that the post FO
distributions may become anisotropic \cite{Magas:2003yp,MC0607,NW_layer}
even for FO hypersurfaces with timelike normal
[in short: {\it timelike surface}],
if the normal, $ d\hat{\sigma}^\mu $, and the velocity four-vector,
$u^\mu$, are (very) different.

These studies led to another FO description, where the initial stages of
the collision with strongly interacting matter were described by
fluid dynamics, while the final, outer spacetime domain (or later times)
was described by weakly interacting particle (and string) transport models,
where the final FO was inherently included, as each particle was tracked,
until its last interaction.
It is important to mention, that in these approaches,
the transition from the FD stage to the molecular dynamics (MD) or cascade
stage happens when the matter crosses a spacetime hypersurface, thus
the conservations laws \cite{Cs87,Cs94} have to be satisfied and the post
FO particle phase space distributions \cite{Bu96,Ta07} have to be used
when the post FO distributions become anisotropic.

In this work for the first time we present a simple covariant solution for
the transition problem and conservation laws for the situations when the
matter after the front is in thermal equilibrium (i.e. it has isotropic
phase space distribution) and has an EoS, but the matter before the
front must not be in an equilibrium state.

Then we discuss the situation where microscopic models are appended
to the fluid dynamical model, which are in, or close to thermal
equilibrium, but the EoS, is not necessarily known.

Subsequently, we present the way to generalize the problem to
anisotropic matter in final state, which is necessary for
FO across spacelike surfaces and also for timelike surfaces
if the flow velocity is large in the rest frame of the front (RFF).
This problem was solved in kinetic approach for the
Bugaev cut-J\"{u}ttner approach \cite{AC99-3309,AL99-388} and
the Csernai-Tamosiunas cancelling-J\"{u}ttner approach,
\cite{Ta07} by calculating the
energy momentum tensors explicitly from the anisotropic phase space
distributions, but no general solution is given for post FO matter
with anisotropic pressure tensor.

\section{Numerical extraction of a Freeze Out hypersurface}

The transition hypersurface between two stages of a dynamical development
are most frequently postulated, governed by the requirement of simplicity.
Thus, such a hypersurface is frequently chosen as a fixed coordinate
time in a descartian frame $t$, or at a fixed proper time $\tau$ from
a spacetime point, although in a general 3+1 dimensional system the choice
of such a point is not uniquely defined. It is important that the {\it
transition hypersurface should be continuous}, (without holes where conserved
particles or energy or momentum could escape through, without being
accounted for). To secure that one quantity (e.g. baryon charge) does
not escape through the holes of a hypersurface is not sufficient,
as other quantities may (e.g. momentum in case if $P dV$ is different
on the two sides of a hole).  Again, to construct such a continuous
hypersurface in a general 3+1 dimensional system is a rather complex task,
although, in 1+1 or 2+1 dimensions it seems to be easy.

Both the initial state models and the intermediate stage, fluid dynamical
models may be such that the calculation could be continued beyond the
point where a transition takes place. Then spacetime location of
the transition to the next stage can or should be decided, based on
a physical condition or requirement, which may be external to the
development itself. As a consequence, in some cases the determination of
transition surface may be an iterative process.

Numerically, the extraction of a Freeze Out (FO) hypersurface is by no means
trivial.
One of us, BRS, has recently provided a proper numerical treatment regarding
the extraction of FO hypersurfaces in two (2D), three (3D) and four (4D)
dimensions~\cite{BRS09,BRS03,BRS04,BRS10}.

\begin{figure*}[t]
\epsfig{width=17.5cm,figure=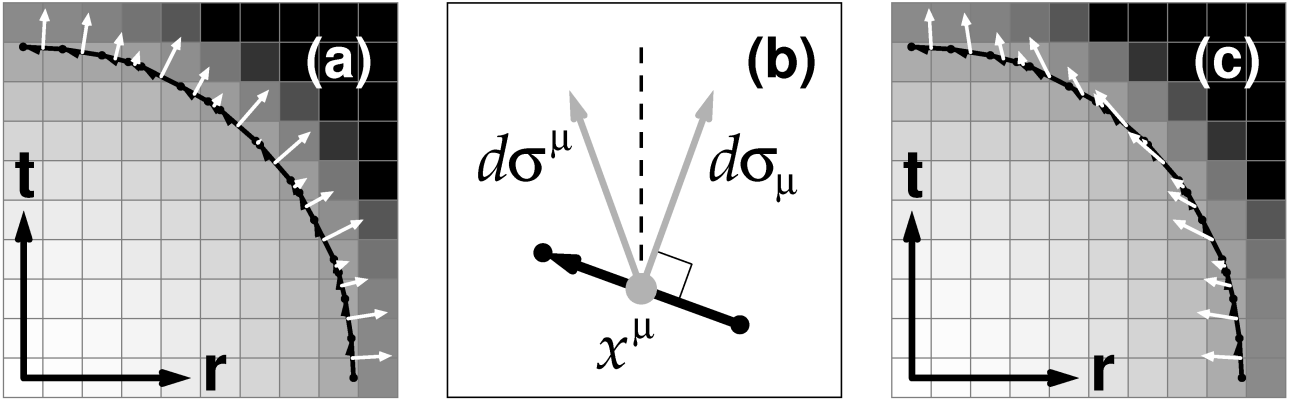}

%\begin{widetext}
%\begin{center}
%\begin{figure*}
%\centerline{\hspace{-0.3in}
%\includegraphics[width=7.0in]{figure1.eps}}
\caption{
(a) A gray-level image representing the temperature evolution
in spacetime of a 1D relativistic fluid (see text) which is superimposed with
contra-variant FO contour vectors (black) and with corresponding
co-variant normal vectors $d\sigma_\mu(x^\mu)$ (white);
(b) co-variant and contra-variant normal vectors, $d\sigma_\mu(x^\mu)$
and $d\sigma^\mu(x^\mu) $, (gray) respectively, which originate at the
contra-variant center $x^\mu$ (gray) of a single contra-variant FO contour
vector (black);
(c) as in (a), but with contra-variant normal vectors $d\sigma^\mu(x^\mu)$
(white) instead of co-variant ones.
} \label{multib_1}
\end{figure*}
%\end{center}
%\end{widetext}

For instance, in 2D the history, i.e., the temporal evolution, of a
temperature field of a one-dimensional (1D) relativistic fluid can be
represented by a gray-level image ({\it cf.}, Fig.~1).
In the figure, we use the time $t$ and the radius $r$ for the temporal and
the spatial dimensions, respectively.
Let bright pixels (i.e., picture elements) refer to high temperatures and
dark ones to low temperatures of the fluid.
In this example, a 2D freeze-out hypersurface is an iso-therme.

In Fig.~1.a, we also depict the corresponding co-variant normal vectors
$d\sigma_\mu(x^\mu)$.
In 2D, the length of each normal vector is equal to the length of each
supporting iso-contour vector.
Each normal vector has its origin at the contra-variant center, $x^\mu$, of a
given contra-variant iso-contour vector and points to the exterior of the
enclosed spacetime region.
The latter is also indicated in Fig.~1.b, where we show that a contra-variant
normal vector $d\sigma^\mu(x^\mu)$ can be obtained by reflection of the
co-variant normal vector $d\sigma_\mu(x^\mu)$ at the time axis (dashed line).

Finally, in Fig.~1.c we show the contra-variant FO contour vectors with their
corresponding contra-variant normal vectors $d\sigma^\mu(x^\mu)$.
Not all of these contra-variant normal vectors point to the exterior of the
enclosed spacetime region.

Note that the sign conventions of the normals of the transition hypersurface
are important, and must be discussed, especially if both timelike and
spacelike surfaces are studied.
In fact, only the timelike contra-variant normal vectors point outwards,
whereas the spacelike contra-variant normal vectors point inwards
\cite{BRSnote}.

If we know the FO hypersurface and the local momentum distribution after
the transition the total, measurable momentum distribution can be
evaluated by the Cooper-Frye formula  \cite{CF74}.

\section{Equations for parameters of final matter in equilibrium}

\indent
Let us define the contra-variant and co-variant surface normal four-vectors as
\begin{eqnarray*}
d\sigma^\mu =& (\sigma_t, \sigma_x, \sigma_y, \sigma_z)
= &(\sigma_t, \vec{\sigma} ) \\
d\sigma_\mu =& (\sigma_t, -\sigma_x, -\sigma_y, -\sigma_z)
= & (\sigma_t, - \vec{\sigma} ) \ ,
\end{eqnarray*}
where in general $d\sigma^\mu d\sigma_\mu = \pm D^2$, as the surface element
can be either spacelike (-) or timelike (+). We can also introduce a
unit normal to the surface as:
$
d\hat{\sigma}^\mu \equiv d\sigma^\mu / D
$
so that
$
d\hat{\sigma}^\mu d\hat{\sigma}_\mu = \pm 1 \; .
$
Furthermore
$
d\hat{\sigma}^\mu = \gamma_\sigma (1, s_x, s_y, s_z)
= \gamma_\sigma (1, \vec{s} ) \
%\ {\rm and} \ \
%d\hat{\sigma}_\mu = \gamma_\sigma (1, -s_x, -s_y, -s_z)\;
%= \gamma_\sigma (1, - \vec{s} ) \ ,
$
where for timelike surfaces $\gamma_\sigma^2 = 1 / (1 - \vec{s}\;^2)$
and for spacelike surfaces $\gamma_\sigma^2 = 1 / (\vec{s}\;^2 - 1)$.
For the frequently used timelike, one-dimensional case
$
d\hat{\sigma}^\mu = \gamma_\sigma (1, \vec{s}) = \gamma_\sigma (1, 0, 0, dt/dz)
$.
\medskip

In the general case the conserved energy-momentum current crossing
the surface element is
\begin{equation}
A^\mu \ = \ T^{\mu\nu} d\sigma_\nu\; \ = \
w\: u^\mu u^\nu\; d\sigma_\nu - P\; g^{\mu\nu}\; d\sigma_\nu \
\label{EM-curr}
\end{equation}
$A^\mu$ must be continuous across the freeze-out surface,
as must the baryon current $N^\mu d\sigma_\mu$,
\begin{equation}
j  =
N^\mu d\sigma_\mu =
n \: u^\mu d\sigma_\mu =
n \: \gamma\: \left( \sigma_t - (\vec{v}\cdot\vec{\sigma}) \right) \ ,
\label{eqN}
\end{equation}
where $j$ is the invariant scalar baryon charge current.

We assume that the initial state, "$0$", and its energy momentum tensor
and baryon current before the front is known.  We aim for the characteristics
of the final state.
In total there are six unknowns in the equilibrated final state, these are
$\vec{v}$, $e$, $P$ and $n$ (here we drop the index "$1$" for the final
state for shorter notation), however the pressure $P$,  a function of $e$
and $n$, is given by the
EoS, $P=P(e,n)$. Knowing $n$ and $e$, the EoS,  and the particular form of the
corresponding equilibrated distribution function, the parameters $T$,
and $\mu$, can also be obtained.

Thus, we have to solve 5 equations:
\begin{eqnarray}
j_{0} &=&
n \;\gamma \;(\sigma_t - (\vec{v}\cdot\vec{\sigma}))
\label{eq9}\\
A_{0t} &=&
w \gamma^2\;(\sigma_t - (\vec{v}\cdot\vec{\sigma}))
- P \: \sigma_t \label{eq0}\\
A_{0x} &=&
w \gamma^2\;(\sigma_t - (\vec{v}\cdot\vec{\sigma})) v_x \;
- P \:\sigma_x \label{eq1}\\
A_{0y} &=&
w \gamma^2\;(\sigma_t - (\vec{v}\cdot\vec{\sigma})) v_y \;
- P \: \sigma_y \label{eq2}\\
A_{0z} &=&
w \gamma^2\;(\sigma_t - (\vec{v}\cdot\vec{\sigma})) v_z \;
- P \: \sigma_z \label{eq3}\;
\end{eqnarray}
The l.h.s. represents quantities of the initial state of matter and
the corresponding conserved quantities are known.
Equations (\ref{eq0},\ref{eq1})
can be solved for $\gamma^2$ in the calculational frame:
\begin{equation}
\gamma^2 = \frac {A_{0t}+P \sigma _t}
{w\; (\sigma_t - (\vec{v}\cdot\vec{\sigma}))}  \ , \ \
\gamma^2 = \frac {A_{0x}+P \sigma _x}
{w\;v_x (\sigma_t - (\vec{v}\cdot\vec{\sigma}))}
\label{eqg2a}.
\end{equation}
Using now Eq. (\ref{eq1},\ref{eq2},\ref{eq3}) one obtains
$ v_x$, and in a similar fashion
$ v_y$ and $ v_z$

\begin{equation}
v_x = \frac{A_{0x}+P \sigma_x}{A_{0t}+P \sigma_t}\ , \ \
v_y = \frac{A_{0y}+P \sigma_y}{A_{0t}+P \sigma_t}\ , \ \
v_z = \frac{A_{0z}+P \sigma_z}{A_{0t}+P \sigma_t}\
\label{v_star}.
\end{equation}
This results for $\gamma^2 = 1 / ( 1- \vec{v}^2 )$, in
\begin{equation}
\gamma^2 = \frac{\left(A_{0t}+P \;\sigma_t\right)^2}
{( A_0^\mu + P\; d\sigma^\mu )^2 }\,
\label{eqg2b}\;.
\end{equation}
where,
$
( A_0^\mu + P\; d\sigma^\mu )^2 =
\left(A_{0t}+P \:\sigma_t\right)^2 -
\left(A_{0x}+P \;\sigma_x\right)^2 -
\left(A_{0y}+P \;\sigma_y\right)^2 -
\left(A_{0z}+P \;\sigma_z\right)^2  \ ,
$
is an invariant scalar, and $\gamma$ transforms as the 0-th component
of the 4-vector $A_0^\mu + P\; d\sigma^\mu$.  Notice that eq. (\ref{eq9}) was
not used up to this point, thus we can use there results both for
the baryon-free and baryon-rich case.

We can have an elegant direct solution for the proper energy density,
$e$, and pressure, $P$, as both of these quantities are invariant
scalars, and we can express these by the covariant,
4-vector equation (\ref{EM-curr}). From this 4-vector equation we can get
two invariant scalar equations by (i) taking its norm, $A_0^\mu A_{0\mu}$,
and (ii) taking its projection to the normal direction,
$A_0^\mu d\sigma_\mu$:
\begin{eqnarray}
A_0^\mu A_{0\mu} & = &
w^2 (u^\mu d\sigma_\mu )^2 + P^2 (d\sigma^\mu d\sigma_\mu ) \nonumber\\
                 & - &
2Pw (u^\mu d\sigma_\mu )u_\mu g^{\mu\nu} d\sigma_\nu
\nonumber\\
                 & = &
w (e-P) (u^\mu d\sigma_\mu )^2 + P^2 (d\sigma^\mu d\sigma_\mu ) \ ,
\label{AA} \\
A_0^\mu d\sigma_\mu & = &
w (u^\mu d\sigma_\mu )^2 - P (d\sigma^\mu d\sigma_\mu ) \ .
\label{Ads}
\end{eqnarray}
Now expressing $ w (u^\mu d\sigma_\mu )^2 $ from eq. (\ref{Ads}) and
inserting it to eq. (\ref{AA}) , we obtain our final equation
\begin{equation}
A_0^\mu A_{0\mu}  =
(e-P) A_0^\mu d\sigma_\mu + e\;P\; (d\sigma^\mu d\sigma_\mu ) \ ,
\label{AA2}
\end{equation}
which
can be solved straightforwardly if the EoS, $P = P(n,e)$, is known. The
other three elements of the equation,
$A_0^\mu A_{0\mu}$,
$A_0^\mu d\sigma_\mu$, and
$d\sigma^\mu d\sigma_\mu$,
are known from the normal to the surface and from energy-momentum current
from the pre-transition side.

Then, eqs. (\ref{eqg2a}-\ref{eqg2b}) can be used to determine the
final flow velocity.
At the end, after all conservation law equations are solved,
we have to check the non-decreasing entropy condition (\ref{entropy})
to see whether the solution is physically possible.
If the overall entropy is decreasing after transition that would mean that
the hypersurface is chosen incorrectly.
One will need to choose more realistic condition for the transition and
repeat the calculations.

This result can be used both if the initial state is in equilibrium and if
it is not.

\subsection{Final Matter with zero Baryon charge}

In case of an ideal gas of massless particles after the front,
with an EoS of $P = e/3$,
eq. (\ref{AA2}) leads to a quadratic equation,
$$
d\hat{\sigma}^\mu d\hat{\sigma}_\mu\, e^2 +
2\, a^\mu d\hat{\sigma}_\mu\, e -
3\, a^\mu a_\mu = 0 \, ,
$$
where $ a^\mu \equiv A_0^\mu / D$, is the energy momentum transfer
4-vector across a unit hypersurface element.

If the flow velocity
is normal to the FO hypersurface, $u^\mu = d\hat{\sigma}^\mu$, then
for an initial perfect fluid in the Local Rest (LR) frame the above
covariant equation takes a simple form,
$$
e^2 + 2\, e_{0}\, e - 3\, e_{0}^2 = 0 \ .
$$
This has two real
roots, $e = e_{0}$ (energy density is conserved) and $ e = - 3 e_{0}$
which does not correspond to a physical solution, as the energy density should
not be negative.
\subsection{Final Matter with Finite Baryon Charge}

If the EoS depends on the conserved baryon charge density also,
then we must exploit in addition eq. (\ref{eqN}):
$$
j_0 \equiv j = n (u^\mu d\sigma_\mu )
$$
and inserting $u^\mu d\sigma_\mu = j/n$ from here to eq. (\ref{Ads}) yields
$$
j^2 \frac{w}{n^2} = A_0^\mu d\sigma_\mu + P (d\sigma^\mu d\sigma_\mu ) \ ,
$$
where ${w}/ {n^2}=X$ is the generalized specific volume, well known from
relativistic shock and detonation theory~\cite{Cs94}.
This equation provides another equation for $e + P$ as
\begin{equation}
\frac{e+P}{n^2} = \frac{1}{j^2} \ \left[
A_0^\mu d\sigma_\mu + P (d\sigma^\mu d\sigma_\mu ) \right] \ ,
\label{eP2nd}
\end{equation}
which, together with eq. (\ref{AA2}) and the EoS, $P=P(e,n)$,
provide three equations to be solved for $e, P$ and $n$.

This evaluation of the post FO configuration is in agreement with
the theory of relativistic shocks and detonations~\cite{Cs87,Ta48}
allowing for both spacelike and timelike FO hypersurfaces.
See also \cite{Cs94}. This method of evaluation observables is
frequently used at the end of fluid dynamical model calculations
(see e.g.\cite{Bravina, Cs2009,Cs2010}).

\section{Transition to Molecular Dynamics before Freeze Out}

Recently a frequently practiced method to describe the final stages
of a reaction is to switch the FD model over to a Molecular Dynamics (MD)
description at a transition hypersurface. This is frequently  a fixed
time, $t$, or fixed proper time, $\tau$ hypersurface.
The generation of the initial state of such an MD model
is a task, which depends on the constituents of the matter described by
the MD model. Nevertheless, same principles must be satisfied, like the
conservation laws, eqs.(\ref{nconserve}-\ref{tconserve}).

\subsection{Equilibrium and EoS known before and after the transition}

Let us assume, although not required by physical laws, that we have
thermal equilibrium on both sides of the transition and we know explicitly
the corresponding final momentum distribution of particles.
Then, the fundamental equation to construct the post transition
microscopic state, in addition to the
conservation laws is the Cooper-Frye formula,
\begin{equation}
E \frac{dN_i}{d^3p} = \int_\sigma \, f_i(x,p) \, p^\mu d\sigma_\mu \, ,
\label{CF-f}
\end{equation}
assuming that the local phase space distribution, $f_i(x,p)$, is known
for all initial components of the MD model. If $f_i(x,p)$ are local
equilibrium distributions then (in principle) we know the intensive and
extensive thermodynamical parameters and the EoS of the matter when
the MD model simulation starts. These must not be the same as the ones
before the transition hypersurface.

In the usual transition from FD to MD models, where the initial state
of MD is in equilibrium, the EoS-s are known on both sides of the transition
surface, and thus, both the equations of Rayleigh-line and
Taub-adiabat, eqs. (\ref{rayligh-taub}), as well as the invariant scalar
equations derived here, eqs.(\ref{AA},\ref{Ads},\ref{AA2},\ref{eP2nd})
can be used to
determine all parameters of the matter starting the MD simulation. These then
determine the phase space distributions, $f_i(x,p)$ of all components of
the MD simulation. Subsequently eq.(\ref{CF-f}) can be used to generate
randomly the initial constituents of the MD simulation.

As  eq.(\ref{CF-f}) is a covariant
equation applicable in any frame of reference,
the most straightforward is to perform the generation of particles in the
calculational frame of the MD model. This transition is by now performed in many hybrid models combining fluid dynamics with microscopic transport models \cite{BD1999}. These models at present are the most effective to describe experimental data and make the need for a Modified Boltzmann Transport Equation \cite{NW_layer} less problematic. 

%The particular steps of the generation of particles may vary on the type of the MD model used, a good example is given in ref. \cite{PS08}.

In some cases the first step of the transition, the determination of the
parameters of the final state from the exact conservation laws, is dropped
with the argument that both before and after the transition the matter
has the same constituents and the same EoS, thus the all extensive and
intensive thermodynamical parameters as well as the flow velocity must
remain the same. Then, using the intensive parameters the final particle
distributions in the Cooper-Frye formula, eq.(\ref{CF-f}), can be directly
evaluated in a straightforward way.   This procedure is correct, but only
if all features of the two states of the matter and their EoS are identical.
In some cases the pre transition EoS assumes effective hadron
masses depending on the matter density, while the final EoS is that of
a hadron ideal gas mixture, but with fixed vacuum masses. This leads to
a difference in the EoS, thus the above procedure is approximate.
In such cases, the method can be used, but the accurate
conservation laws can be enforced by a final adjustment step described in
the next subsection.

The situation is similar if the constituents and the EoS are almost identical
before and after the transition, but before the transition a weak or
weakening mean field potential or compression energy is taken into
account.

\subsection{Enforcing Conservation Laws with Approximate
            Generation of the Final State}

In addition to the above mentioned approximate methods, even for really
identical EoS-s across the transition or with generating the final
EoS parameter based on conservation laws for the final EoS, inaccuracies
may arise due to other reasons:
during the random generation of the initial constituent particles of the
MD simulation, the exact conservation laws
may be violated, due to finite number effects.
However, the energy and particle number
conservations are usually enforced during the random generation of particles,
even if the above procedure of solving the conservation laws beforehand is
not fully followed.  This is usually the consequence of the fact that the
EoS of the MD model is not necessarily known if the model has complex
constituents and laws of motion.

In any case to remedy this random error and make the conservation laws
exactly satisfied a final correction step is advisable, and it is not
always performed.
If the energy and particle number conservations are enforced
then, the last variable to balance is the momentum conservation.
This regulates the flow velocity of the matter after the transition initiating
the MD simulation.

The energy momentum tensor and baryon current for the generated
random set of particle species, "$i$', for
each fluid cell (or group of cells if the multiplicity in a single cell
is too low) can be calculated from the kinetic definition:
\begin{eqnarray*}
T^{\mu\nu}(x) &=& \sum_i \int \frac{d^3p_i}{p_i^0} p_i^\mu p_i^\nu f_i(x,p_i),\\
N^{\mu}(x)    &=& \sum_i \int \frac{d^3p_i}{p_i^0} p_i^\mu f_i(x,p_i).
%\label{emtkin}
\end{eqnarray*}
which, yield the resulting momentum and flow velocity of the
matter. This can be used to adjust the flow velocity to
achieve exact conservation of momentum, and modify the velocity of
generated particles by the required Lorentz boost.
The other conserved quantities may then be affected also,
but an iterative procedure to eliminate the error completely
is not crucial as the error can be given quantitatively.

If the randomly generated state is not following a thermal equilibrium
phase space distribution, $f_i(x,p_i)$, and thus does not have an EoS, the
above described scalar equations cannot be used to generate the initial
configuration of the MD model. Nevertheless, the second step to check the
conservation laws with the kinetic definition, and then correct the
parameters of the generated particles can be done. For a required level of
accuracy in this case an iterative procedure may be necessary.

Another, easier way to remedy this problem is to choose the transition
hypersurface earlier so that the subsequent matter is still in thermal
equilibrium. This can always be done if the requirement of entropy increase
is satisfied.

\section{Final state out of thermal equilibrium}

We have mentioned that the assumption for having thermal equilibrium in the
final state is neither excluded nor required from transport theoretical
considerations. However, thermal equilibrium distribution is not possible
if we have to describe FO across a spacelike hypersurface (see the
discussion in section \ref{intro}.)

In the MD model description the final post FO momentum distributions develope
a local anisotropy if the FO has locally a preferred direction.
Unless the unit normal of the FO hypersurface is equal to the local flow velocity
of the pre FO matter, there is always a selected spatial direction which is the
dominant direction of FO.  This situation is discussed in several
theoretical works, and some general features can be extracted from
these studies.

\subsection{Approximate kinetic models for Freeze Out}

In explicit transport models this situation is handled
\cite{Bu96,AC99-3309,Ta07,AL99-388}: starting from an equilibrium J\"{u}ttner
distribution and considering a momentum dependent escape probability in the
collision term, - which reflected the direction of the FO front and
the distance from the front, - an anisotropic distribution was obtained
(i.e. a distribution, which was anisotropic even in its own LR frame).

This anisotropic distribution could be approximated
with analytic distribution functions \cite{Bu96,AC99-3309}:
The starting point is the un-cut, isotropic, J\"{u}ttner distribution in the
rest frame of the gas (RFG), which is
centered around the 4-velocity vector, $u^\mu_{RFG}$. This distribution
is then cut or cut and smoothed. The resulting distribution has a different
new flow velocity,  $u^\mu_{LR}$, which is non-zero in
RFG, and is pointing in space in the direction of the normal
of the FO hypersurface, $\vec{\sigma}$, labeled by $\parallel$.
This $u^\mu_{LR}$ defines the Local Rest (LR) frame
of the post FO matter.

The spatial direction of $\vec{\sigma}$ is not affected by the Lorentz
transformation from RFF to RFG and then to LR, as $\vec{\sigma}$ is the
direction of the Lorentz transformation from RFG to LR
\footnote{In general the spatial components of the FO surface, $\vec{\sigma}$,
must not be parallel to $\vec{v}_{RFG}$ or $\vec{v}_{LR}$ but these latter
velocities can be decomposed to $\parallel$ and $\perp$ components with
respect to $\vec{\sigma}$. Due to the construction
\cite{Bu96,AC99-3309,Ta07} of the cut- or cancelling-J\"{u}ttner distributions
$(\vec{v}_{LR} - \vec{v}_{RFG}) \parallel \vec{\sigma}$ or
$\vec{v}_{LR,\perp} = \vec{v}_{RFG,\perp}$.}.
In the general case the boost in the $\vec{v}_{RFG,\perp}$
direction leads to a change
of the distribution function in the $\vec{p}_\perp$ direction, but
does not affect the distribution in the $\vec{p}_\parallel$ direction, or
the procedure of cutting or cancelling the distribution in the
$\parallel$ direction. (The illustration in Fig. 2a shows the
spatial momentum distribution where the boost in the orthogonal direction
$\vec{v}_{RFG,\perp}$ is already performed.)

In the final LR frame, the matter is characterized
by a rather complex energy momentum tensor, inheriting some parameters
from the original uncut distribution in RFG, like the temperature and
chemical potential, but as the resulting distribution is not a
thermal equilibrium distribution, these parameters are not
playing any thermodynamical role. One has to determine all parameters
numerically from conservation laws (\ref{nconserve},\ref{tconserve}),
as done in refs. \cite{AC99-3309,Ta07}.

Interestingly, a simplified numerical kinetic FO model \cite{AL99-388}
led to a FO distribution satisfying the condition
$p^\mu d\sigma_\mu > 0$ for spacelike FO with a smooth distribution
function, which is anisotropic (also in its own LR frame) and has a
symmetry axis pointing in the dominant FO direction. This distribution
was then approximated with an analytic, "cancelling-J\"{u}ttner" distribution
\cite{Ta07}, which can also be used to solve the FO problem.

After FO, the symmetry properties of the energy momentum tensor
are the same for the cut-J\"{u}ttner and cancelling-J\"{u}ttner cases
\cite{Bu96,AC99-3309,Ta07}.
The FO leads to an anisotropic
momentum distribution and therefore to an anisotropic pressure
tensor.  The energy momentum tensor is not
diagonal in the RFG frame, there is a
non vanishing transport term, $T^{0i}$
\cite{AC99-3309,Ta07},
in the 2-dimensional plane spanned by the  4-vectors,
$u^\mu_{RFG}$ and $d\hat{\sigma}_\mu$.
One can, however, diagonalize the energy momentum tensor by making a Lorentz
boost into the LR frame using Landau's definition for
the 4-velocity, $u^\mu_{LR}$. In this frame
then the energy momentum tensor becomes diagonal, but the pressure terms
are not identical, due to the anisotropy of the distribution:
\begin{equation}
T^{\mu\nu}=\left.
{\rm diag}(e, P_\parallel, P_\perp, P_\perp )
\phantom{\int} \right| _{LR}
\label{Tdiag}
\end{equation}
Here the energy density, $e$, of course must not be the same as in the
case of an isotropic, thermal equilibrium post FO momentum distribution.
This can be seen from the kinetic definition of the energy momentum tensor
as shown in refs. \cite{AC99-3309,Ta07}.

We need
the complete post FO momentum distribution and the corresponding
energy momentum tensor to determine final observables. This depends on the
transport processes at FO, and cannot be given in general; however,
due to the symmetries of the collision integral, the symmetries
of the energy momentum tensor are the same irrespectively of the
ansatz used (e.g. cut-J\"{u}ttner, cancelling-J\"{u}ttner or some other
distribution).

\begin{figure*}[t]
\epsfig{width=8.3cm,figure=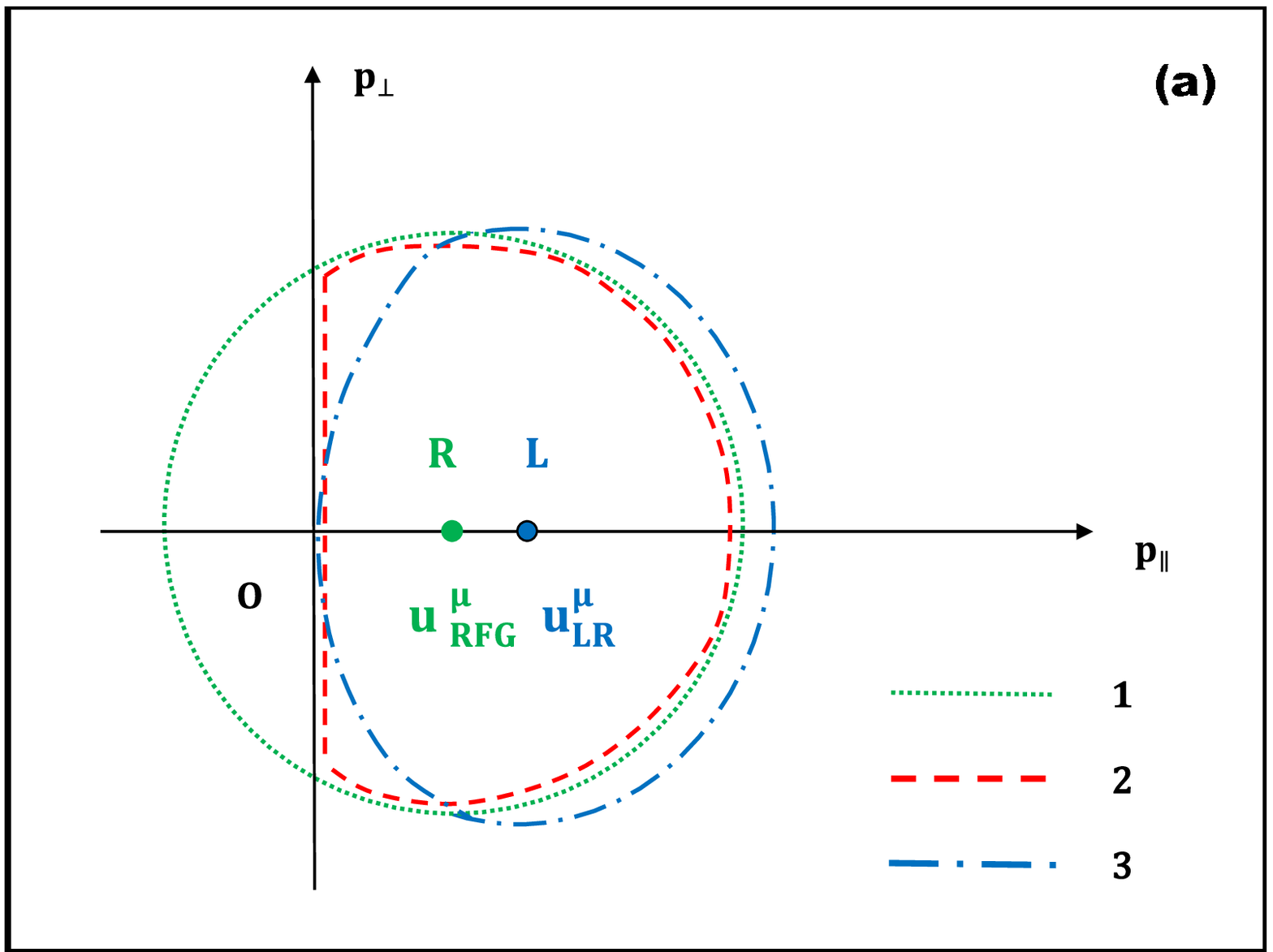}
\epsfig{width=8.6cm,figure=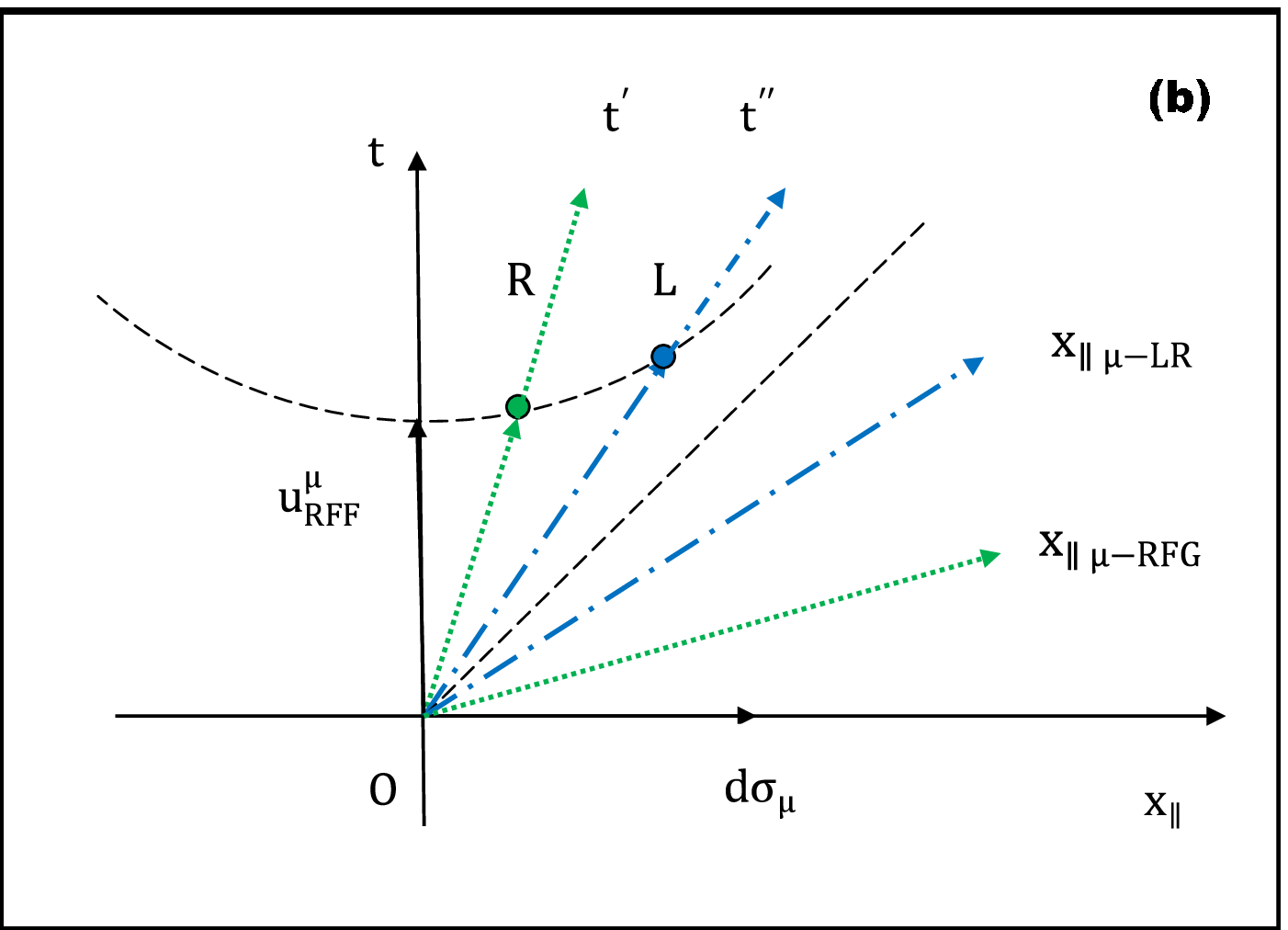}
%\begin{widetext}
%\begin{center}
%\begin{figure}
%\centerline{\hspace{-0.3in}
%      \includegraphics[height=6cm,width=8.5cm]{fa.eps}
%      \includegraphics[height=6cm,width=8.5cm]{fb.eps}}
\caption{
(Color online)
(a)  Illustrative contour plot of momentum distributions of
particles after freeze-out (FO) to an anisotropic final state. Here the
$p_{\parallel}$-axis points into the direction of the space-like normal of
the FO hypersurface, while $p_{\perp}$ is the orthogonal direction to
$p_{\parallel}$. Line $"1"$ indicates the un-cut J\"{u}ttner distribution
(dotted-line), $"2"$ indicates the cut-J\"{u}ttner distribution (dashed-line),
$"3"$ indicates the canceling-J\"{u}ttner distribution (dashed-dotted line).
$"R "$ is the center of the spherical un-cut J\"{u}ttner distribution moving
with velocity, $u_{RFG}^{\mu}$, and $"L"$ is the center of the cut and
canceling-J\"{u}ttner distributions, which move to the right along the
parallel momentum direction with velocity, $u_{LR}^{\mu}$, corresponding to
their LR frame. If $u_{RFG}^{\mu}$ and $u_{LR}^\mu$ have a non-vanishing
$\perp$ component the resulting distribution can be obtained by an additional
Lorentz boost in that direction.
(b) Space-time reference frames for anisotropic final state distributions.
The time axis, $t$, and the spatial axis, $x_{\parallel}$, represent the
"Rest Frame of Front" (RFF), which is the defining front for the
cut-J\"{u}ttner and canceling-J\"{u}ttner distributions. $"L"$ indicates the
"Local Rest Frame" (LR) of the post FO cut or canceling-J\"{u}ttner
distribution. $"R"$ indicates the "Rest Frame of Gas" (RFG) corresponding to
the isotropic, un-cut J\"{u}ttner distribution. Thus the velocities
$u_{RFF}^{\mu}$, $u_{RFG}^{\mu}$ and  $u_{LR}^{\mu}$ point in the directions
of the time axes of the corresponding reference frames, indicated by
$t$, $t^{'}$, $t^{''}$ respectively. $d\sigma_{\mu}$ is the normal vector of
the front, it has the same direction as the spatial axis, $x_{\parallel}$.
The spatial axes of the frames RFG (R) and LR (L) are shown with the same
line style (color) as the corresponding distributions. The LR frames for case
2 and 3 are the same.
}
\label{fab}
\end{figure*}

In kinetic transport approaches the microscopic escape
probability \cite{MC0607} is peaking in the direction of
$d\hat{\sigma}_\mu$, which yields a distribution peaking in
this direction, i.e. yielding the same symmetry properties as
the previously mentioned analytic ansatzes.
The energy momentum tensor in general takes the form
\begin{equation}
T^{\mu\nu}=e\, u_{LR}^{\mu}u_{LR}^{\nu}- P_\perp \Delta_{LR}^{\mu\nu}
       +  (P_\parallel - P_\perp) \hat{F}^\mu \hat{F}^\nu
\label{TSmunu}
\end{equation}
where
$\Delta_{LR}^{\mu\nu}$ is the orthogonal projector to $u_{LR}^{\mu}$,
and
$\hat{F}_\mu$ is the unit 4-vector projection of $d\sigma_\mu$
in the direction orthogonal to  $u_{LR}^{\mu}$, i.e.
$\hat{F}^\mu= C \Delta_{LR}^{\mu\nu} d\sigma_\nu$, where $C$ ensures
normalization to -1. In the Landau LR frame
this returns expression (\ref{Tdiag}).
The 4-velocity, $u_{LR}^{\mu}$, and the other parameters of the
post FO state of matter, should be determined from the
conservation laws (\ref{nconserve},\ref{tconserve}).
The schematic diagram of the asymmetric distributions
and the different reference frames can be seen in Fig. \ref{fab}.

%\end{center}
%\end{widetext}

The FO problem was solved for these configurations and ansatzes,
by satisfying the conservation laws explicitly for the
full energy momentum tensor.
We do not have a general EoS(s) that would characterize the connection among
$e$, $P_\parallel$ and $P_\perp$, furthermore the relation connecting
these quantities depends on the 4-vectors
$ d\hat{\sigma}^\mu$ and $u_{LR}^\mu$. In addition this
connection depends on the details or assumptions of the transport
model. The simple models  \cite{AC99-3309,Ta07} provide examples
for such a dependence. If $d\hat{\sigma}^\mu$ is known, then
for baryon free matter we can determine four unknowns:  $u_{LR}^\mu$ and,
an additional parameter of the post FO distribution,
from eq. (\ref{tconserve}).
(Due to normalization only 3 components of
$u_{LR}^\mu$ are unknowns.) For baryon-rich matter we can determine one
more unknown parameter, since we have one additional equation, the
conservation of baryon charge from eq. (\ref{nconserve}).

\subsection{Exploiting general symmetries of anisotropic final states}

The first step of solution can be done similarly to the
isotropic case.
Then in eq.  (\ref{EM-curr}) the enthalpy will change as
$w \rightarrow  e + P_\perp \equiv w_\perp$, and
$P \rightarrow  P_\perp$, plus an additive
term will appear,
$ (P_\parallel - P_\perp)\; \hat{F}^\mu  \hat{F}^\nu\; d\sigma_\nu$.
Furthermore, eqs. (\ref{eq0}-\ref{eq3})
remain of the same form, with $w_\perp$, and
$P_\perp$, plus the additive term
$ (P_\parallel - P_\perp)\; \hat{F}^\mu  \hat{F}^\nu\; d\sigma_\nu$ will appear
in the r.h.s. of eqs.(\ref{eq0}-\ref{eq3}).
This additive term will also appear
in the expression of $v_x$ after eq. (\ref{eqg2a}) and in the denominator
of eq.  (\ref{eqg2b}) also.

The additional term,
$ (P_\parallel - P_\perp)\; \hat{F}^\mu  \hat{F}^\nu\; d\sigma_\nu$,
 in eq.  (\ref{EM-curr}) is orthogonal to $u^\mu$ (by definition of
$\hat{F}^\mu$), so
when we calculate the scalar product (\ref{AA}) their cross term
vanishes, so
\begin{eqnarray}
A_0^\mu A_{0\mu} & = &
w_\perp (e-P_\perp) (u^\mu d\sigma_\mu )^2 +
(P_\perp)^2 (d\sigma^\mu d\sigma_\mu )
\nonumber \\
& - &
(P_\parallel-P_\perp) (P_\parallel+P_\perp) (\hat{F}^\nu\; d\sigma_\nu)^2 \ ,
\label{AA-p} \\
A_0^\mu d\sigma_\mu & = &
w_\perp (u^\mu d\sigma_\mu )^2 - P_\perp (d\sigma^\mu d\sigma_\mu ) \nonumber\\
                    & + &
 (P_\parallel-P_\perp) (\hat{F}^\nu\; d\sigma_\nu)^2 \ .
\label{Ads-p}
\end{eqnarray}
Now one can express $ w_\perp (u^\mu d\sigma_\mu )^2 $ from eq. (\ref{Ads-p})
and inserting it to eq. (\ref{AA-p}), we obtain that
\begin{align}
A_0^\mu A_{0\mu}  =
(e-P_\perp) A_0^\mu d\sigma_\mu +
e\; P_\perp\; (d\sigma^\mu d\sigma_\mu ) \nonumber\\
-(e + P_\parallel) (P_\parallel - P_\perp)  (\hat{F}^\nu\; d\sigma_\nu)^2 \ .
\label{AA2-p}
\end{align}
where this equation is not a scalar equation as it dependes
on $\hat{F}^\mu = C \Delta_{LR}^{\mu\nu} d\sigma_\nu$, where
the projector is dependent on $u^\mu$.
These equations are similar to the ones obtained for the
isotropic case, however, to solve this last equation
we need a more complex relation among
$e$, $P_\parallel$, $P_\perp$. As these arise from
the collision integral in the BTE approach the needed
relation may depend on $u_0^\mu$ and $d\sigma_\mu$.
On the other hand,
the escape probability may be simple, or may be approximated
in a way, which yields an ansatz for this relation with
adjustable parameters, and then
the problem is solvable. This was the case in refs.
\cite{MC0607,NW_layer}.

The recent covariant formulation of the kinetic freeze-out
description \cite{MC0607} indicates that the relation among the
different parameters of the anisotropic energy momentum tensor,
should be possible
to express in terms of invariant scalars, which may facilitate
the solution of the anisotropic FO problem.

When the adjustable parameters of the post FO matter are determined in
this way from the conservation laws, we still need the
underlying anisotropic momentum distribution of the emitted particles
in order to evaluate the final particle spectra using the
Cooper-Frye formula with this anisotropic distribution function.
Once again, when all conservation law equations are solved we have to
check the non-decreasing entropy condition to see whether the
selected FO hypersurface is realistic.

In case of an anisotropic final state, due to the
increased number of parameters and their more involved relations, the
covariant treatment of the problem may not provide a simplification,
compared to the direct solution of conservation laws for each
component of the energy momentum tensor (e.g. \cite{AC99-3309,Ta07}).

\subsection{Anisotropic initial and final states}

Recent viscous fluid dynamical calculations evaluate the anisotropy
of the momentum distribution is in the pre FO viscous flow
(see e.g. \cite{MoHu09}.) This anisotropy is governed by the
spacetime direction of the viscous transport.
The pre and post FO matter may still be different, e.g. the
pre FO state may be viscous QGP with current quarks and
perturbative vacuum, while post FO we may have a hadron gas or
constituent quark gas. The final state will also be anisotropic,
not only because of the initial anisotropy but also due to
freeze-out.
The two physical processes leading to anisotropy are
independent, so their dominant directions are in general different.
In this case the general symmetries are uncorrelated and cannot be
exploited to simplify the description of the transition.
Due to the change of the matter properties, the conservation laws,
eqs. (\ref{nconserve}-\ref{entropy}),
are needed to determine the parameters of the post FO matter before
the Cooper-Frye formula with non-equilibrium post FO distribution
is applied to evaluate observables.

\bigskip

\section{Summary}

In this work a  new simple covariant treatment is presented for
solving the conservation laws across a transition hypersurface.
This leads to a significant simplification of the calculation
if both the initial and final states are in thermal
equilibrium. The same method can also be used for the more
complicated anisotropic final state, however,
this method is only advantageous if the more involved relations
among the parameters of the post FO distribution and the distribution
itself is given in covariant form, preferably through invariant scalars.


\begin{thebibliography}{}
\bibitem{La53}
L.~D.~Landau, Izv. Akad. Nauk SSSR 17 (1953) 51.

\bibitem{Mi5861}
G.~Milekhin, Zh. Eksp. Teor. Fiz. 35 (1958) 1185;
Sov. Phys. - JETP 35 (1959) 829; and
G.~A.~Milekhin, Trudy FIAN 16 (1961) 51.

\bibitem{CF74}
F.~Cooper and G.~Frye, Phys. Rev. D, 10 (1974) 186.

\bibitem{T73}
G.F.~Chapline, M.H.~Johnson, E.~Teller and M.S.~Weiss,
Phys. Rev. D8 (1973) 4302

\bibitem{G74}
W.~Scheid, H.~M\"{u}ller and W.~Greiner,
Phys. Rev. Lett. 32 (1974) 741.

\bibitem{Cs87}
L.~P.~Csernai, Sov. JETP  65 (1987) 216;
Zh. Eksp. Theor. Fiz.  92 (1987) 379.

\bibitem{Cs94}
L.~P.~Csernai: Introduction to Relativistic Heavy Ion Collisions
(Wiley, New York, 1994).

\bibitem{Ta48}
A.~H.~Taub, Phys. Rev.  74 (1948) 328.

\bibitem{Ma0102}
V.K.~Magas, L.P.~Csernai, D.D.~Strottman, Phys. Rev. C64 (2001), 014901, and
V.K.~Magas, L.P.~Csernai, D.D.~Strottman, Nucl. Phys. A 712 (2002) 167.

\bibitem{Bu96}
K.~A.~Bugaev, Nucl. Phys. A  606 (1996) 559.

\bibitem{AC99-3309}
Cs.~Anderlik, L.~P.~Csernai, F.~Grassi, W.~Greiner, Y.~Hama, T.~Kodama,
Zs.~I.~Lazar, V.~K.~Magas, and H.~St\"ocker, Phys. Rev. C 59 (1999) 3309.

\bibitem{Ta07}
K.~Tamosiunas, L.~P.~Csernai. Eur. Phys. J. A20 (2004) 269.

\bibitem{AL99-388}
Cs.~Anderlik  {\it et al.},
Phys. Rev. C 59 (1999) 388;
V.~K.~Magas {\it et al.},
  %``Kinetic freeze-out models,''
  Heavy Ion Phys.\  {\bf 9}, 193 (1999);
   %``Large p(t) enhancement from freeze out,''
  Phys.\ Lett.\  B {\bf 459}, 33 (1999);
  %``Freeze-out in hydrodynamical models in relativistic heavy ion
  %collisions,''
  Nucl.\ Phys.\  A {\bf 661}, 596 (1999).

\bibitem{Magas:2003yp}
  V.~K.~Magas, A.~Anderlik, C.~Anderlik and L.~P.~Csernai,
  %``Non-equilibrated post freeze out distributions,''
  Eur.\ Phys.\ J.\  C {\bf 30}, 255 (2003)

\bibitem{MC0607}
E.~Molnar, L.~P.~Csernai, V.~K.~Magas, A.~Nyiri, K.~Tamosiunas,
Phys. Rev. {\bf C74}, 024907 (2006);  J. Phys. G 34 (2007) 1901;
E.~Molnar, L.~P.~Csernai and V.~K.~Magas,
  %``Covariant kinetic freeze out description through a finite space-time
  %layer,''
Acta Phys.\ Hung.\  A {\bf 27}, 359 (2006);
V.~K.~Magas, L.~P.~Csernai and E.~Molnar,
 %``Freeze out in narrow and wide layers,''
Acta Phys.\ Hung.\  A {\bf 27}, 351 (2006).

\bibitem{NW_layer}
V.~K.~Magas, L.~P.~Csernai and E.~Molnar,
%``Freeze out of the expanding system,''
Eur.\ Phys.\ J.\  A {\bf 31}, 854 (2007);
%``Bjorken expansion with gradual freeze out,''
Int.\ J.\ Mod.\ Phys.\  E {\bf 16}, 1890 (2007);
V.~K.~Magas and L.~P.~Csernai,
%``Kinetic description of particle emission from expanding source,''
Phys.\ Lett.\  B {\bf 663}, 191 (2008);L.P. Csernai, V.K. Magas, E. Molnar et al.,
Eur. Phys. J. C 25, 65 (2005);V.K. Magas, L.P. Csernai, E. Molnar et al.,
Nucl. Phys. A 749, (2005).

%\bibitem{Cs2005c25}
% Modified Boltzmann Transport Equation and Freeze Out
%\bibitem{MagasA749}
% Modified Boltzmann transport equation



\bibitem{Grassi-volFO}
F.~Grassi, Y.~Hama, S.~S.~Padula, and O.~Socolowski,
Phys. Rev. C 62 (2000) 044904.

\bibitem{BRS09}
B.~R.~Schlei, ``A new computational framework for 2D shape-en\-closing
contours,''
Image and Vision Computing 27 (2009) 637, doi: 10.1016/j.imavis.2008.06.014.

\bibitem{BRS03}
B.~R.~Schlei, ``VESTA - Surface Extraction,'' Theoretical Division - Self
Assessment, Special Feature, a portion of LA-UR-03-3000, Los Alamos (2003)
37.

\bibitem{BRS04}
B.~R.~Schlei, ``Hyper-Surface Extraction in Four Dimensions,'' Theoretical
Division - Self Assessment, Special Feature, a portion of LA-UR-04-2143,
Los Alamos (2004) 168.

\bibitem{BRS10}
B.~R.~Schlei, ``Volume-Enclosing Surface Extraction,'' in preparation.

\bibitem{BRSnote}
Note, that we actually use for hypersurface construction in 1+1, 2+1, and
3+1 dimensional numerical simulations the corresponding computer
codes, i.e., DICONEX, VESTA, and STEVE, respectively.
Ref.~\cite{BRS09} explains in great detail the extraction of an oriented
FO contour which is represented by a set of contra-variant (so-called
``DICONEX iso-contour'') vectors.
In 2D, the simplices which represent a hypersurface best are line segments,
whereas in 3D and 4D they are triangles~\cite{BRS03,BRS10} and
tetrahedrons~\cite{BRS04}, respectively.
In particular, the contra-variant 2D FO contour vectors are oriented
counter-clockwise around the enclosed spacetime regions.
The co-variant normals of the contra-variant simplices are obtained from
calculating the mathematical duals of these simplices with respect to a
geometric product ({\it cf.}, e.g., Ref.~\cite{CP09}) within the N-dimensional
multi-linear space under consideration.
Note, that the co-variant normal vectors do not depend on any given metric
tensor, whereas the contra-variant normal vectors do~\cite{BRS10}.

\bibitem{CP09}
C.~Perwass, \textit{Geometric Algebra with Applications in Engineering},
Geometry and Computing, Springer, 2009.

\bibitem{Bravina}
L. Bravina, L.P. Csernai, P. L\'{e}vai, and D. Strottman, PRC 50 (1994) 2161.

\bibitem{Cs2009}
L.P. Csernai, Y. Cheng, Sz. Horv\'{a}t, V. Magas, D. Strottman and
M.Z\'{e}t\'{e}nyi, J. Phys. G 36 (2009) 064032.

\bibitem{Cs2010}
L.P. Csernai, Y. Cheng, V.K. Magas. I.N. Mishustin and D. Strottman,
Nucl. Phys. A 834 (2010) 261c.

\bibitem{BD1999}
% hadronic freeze-out following a first order hadronization phase transition in Ultrarelativistic heavy-ion collisions
S.B. Bass, A. Dumitru, M. Bleicher et al.,
Phys. Rev. C60, 021902 (1999); D. Teaney, J. Lauret and E.V. Shuryak,
Nucl. Phys. A 698, 479 (2002); S.A. Bass, T. Renk, J. Ruppert et al.,
J. Phys. G 34, S979 (2007); C. Nonaka, M. Asakawa and S.A. Bass,
J. Phys. G 35, 104099 (2008); H. Petersen, J. Steinheimer, G. Burau, M. Bleicher, H. St\"ocker,
Phys. Rev. C 78 (2008) 044901; T. Hirano and Y. Nara,
Phys. Rev. C 79, 064904 (2009).

%\bibitem{Tea2002}
% Hydro plus Cascade, flow, the Equation of State, predictions and data


%\bibitem{BR2007}
% Hard and soft probe-medium interactions in a 3D hydro+micro approach at RHIC


%\bibitem{NAB2008}
% The 3D hydro+UrQMD model with the QCD critical point


%\bibitem{PS08}
% Fully integrated transport approach to heavy ion reactions with an intermediate hydrodynamic stage


%\bibitem{HN2009}
% Eccentricity fluctuation effects on elliptic flow in relativistic heavy ion collisions


\bibitem{MoHu09}
P.~Huovinen and D.~Molnar, Phys. Rev. C 79 (2009) 014906.

\end{thebibliography}
\end{document}